\documentclass[12pt]{article}
\usepackage{graphicx}
\usepackage[margin=1.25in]{geometry}
\usepackage[usenames,dvipsnames]{color}
\usepackage{url}
\usepackage[colorlinks = true,
            linkcolor = blue,
            urlcolor  = blue,
            citecolor = blue,
            anchorcolor = blue]{hyperref}

\newcommand\pubnumber{DESY 20-112,\\ KEK Preprint 2020-8,\\ 
 IFIC/20-34, LCTP-20-14\\ SLAC-PUB-17543 }
\newcommand\pubdate{July, 2020}

\def\KEK{High Energy Accelerator Research Organization (KEK), Tsukuba,
  Ibaraki, JAPAN  }
\def\Tokyo{ICEPP, University of Tokyo, Hongo, Bunkyo-ku, Tokyo,
  113-0033, JAPAN}
\def\SNU{Dept. of Physics and Astronomy, Seoul National
  Univ.,  Seoul 08826, KOREA}
\def\DESY{DESY, Notkestrasse 85, 22607 Hamburg, GERMANY}
\def\Berlin{Institut f\"ur Physik, Humboldt-Universit\"at zu Berlin, 12489 Berlin, GERMANY}
\def\SLAC{SLAC,
    Stanford University, Menlo Park, CA 94025, USA}

\def\Peking{Department of Physics, Peking University, Beijing 100871, CHINA}
\def\Osaka{Department of Physics, Osaka University, Machikaneyama, Toyonaka, Osaka 560-0043, JAPAN}
\def\IPMU{Kavli Institute for the Physics and Mathematics of the Universe,
University of Tokyo, Kashiwa 277-8583, JAPAN}
\def\Cornell{Laboratory for Elementary Particle Physics, Cornell
  University, Ithaca, NY 14853, USA}
\def\Orsay{Université Paris-Saclay, CNRS/IN2P3, IJCLab, 91405 Orsay, FRANCE}
\def\Munich{Max-Planck-Institut f\"ur Physik, F\"ohringer Ring 6,
  80805 Munich, GERMANY}
\def\Michigan{Michigan Center for Theoretical Physics, University of Michigan, Ann Arbor,
MI 48109, USA}

\def\Oregon{Department of Physics, University of Oregon, Eugene, Oregon
97403-1274, USA}
\def\Berkeley{ Department of Physics, University of California, Berkeley, CA 94720, USA}
\def\LBNL{Theoretical Physics Group, Lawrence Berkeley National Laboratory, Berkeley,
CA 94720, USA}
\def\Valencia{IFIC (University of Valencia/CSIC), Valencia,  SPAIN}
\def\Kyushu{Department of Physics, Kyushu University, Fukuoka, JAPAN}
\def\Kansas{Department of Physics and Astronomy, University of Kansas,
  Lawrence,  KS 66045, USA}
\def\Sendai{Department of Physics, Tohoku University, Aoba-ku, Sendai
  980-8578,  JAPAN}
\def\Warsaw{Faculty of Physics, University of Warsaw, ul. Pasteura 5,
  02-093 Warszawa, POLAND}
\def\PNNL{Pacific Northwest National Laboratory, Richland, WA 99352,
  USA}
\def\UCSC{Department of Physics, University of California, Santa Cruz,
  CA   95064  USA}

\def\Title#1{\begin{center} {\Large #1 } \end{center}}
\def\Author#1{\begin{center}{ \sc #1} \end{center}}

\newcommand\pubblock{\rightline{\begin{tabular}{l} \pubnumber\\
         \pubdate \end{tabular}}}
\newenvironment{Abstract}{\begin{quotation} \begin{center}
                       ABSTRACT
     \end{center}\bigskip  }{\end{quotation}}

\def\Acknowledgements{\bigskip  \bigskip \begin{center} \begin{large}
             \bf ACKNOWLEDGEMENTS \end{large}\end{center}}

\def\beq{\begin{equation}}
\def\eeq#1{\label{#1}\end{equation}}
\def\eeqn{\end{equation}}

\newenvironment{Eqnarray}%
   {\arraycolsep 0.14em\begin{eqnarray}}{\end{eqnarray}}
\def\beqa{\begin{Eqnarray}}
\def\eeqa#1{\label{#1}\end{Eqnarray}}
\def\eeqan{\end{Eqnarray}}

\let\bar=\overbar

\def\etal{{\it et al.}}

\def\eg{{\it e.g.}}

\def\lsim{\mathrel{\raise.3ex\hbox{$<$\kern-.75em\lower1ex\hbox{$\sim$}}}}
\def\gsim{\mathrel{\raise.3ex\hbox{$>$\kern-.75em\lower1ex\hbox{$\sim$}}}}

\def\half{\frac{1}{2}}

\def\del{\partial}
\def\Dslash{\not{\hbox{\kern-4pt $D$}}}
\def\dslash{\not{\hbox{\kern-2pt $\del$}}}

\def\Dlr{\mathrel{\raise1.5ex\hbox{$\leftrightarrow$\kern-1em\lower1.5ex\hbox{$D$}}}}

\def\ee{e^+e^-}
\def\sstw{\sin^2\theta_w}

\def\MSB{{\bar{M \kern -2pt S}}}
\def\msb{{\bar{\scriptsize M \kern -1pt S}}}

\def\drb{{\bar{\scriptsize D \kern -1pt R}}}

\def\eps{\epsilon}

\def\s#1{\widetilde{#1}}

\makeatletter
\def\section{\@startsection{section}{0}{\z@}{5.5ex plus .5ex minus
 1.5ex}{2.3ex plus .2ex}{\large\bf}}
\def\subsection{\@startsection{subsection}{1}{\z@}{3.5ex plus .5ex minus
 1.5ex}{1.3ex plus .2ex}{\normalsize\bf}}
\def\subsubsection{\@startsection{subsubsection}{2}{\z@}{-3.5ex plus
-1ex minus  -.2ex}{2.3ex plus .2ex}{\normalsize\sl}}

\renewcommand{\@makecaption}[2]{%
   \vskip 10pt
   \setbox\@tempboxa\hbox{\small #1: #2}
   \ifdim \wd\@tempboxa >\hsize     
       \small #1: #2\par          
     \else                        
       \hbox to\hsize{\hfil\box\@tempboxa\hfil}
   \fi}

\makeatother

\begin{document}
\begin{titlepage}
\pubblock

\vfill
\Title{ILC Study Questions for Snowmass 2021}
\bigskip

\bigskip 

\Author{LCC Physics Working Group}
\vfill

\Author{Keisuke Fujii$^1$, Christophe Grojean$^{2,3}$, Michael
  E. Peskin$^4$  (Conveners); Tim Barklow$^4$, Yuanning Gao$^5$,
  Shinya Kanemura$^6$, \\  Jenny List$^2$, Mihoko
  Nojiri$^{1,7}$, Maxim Perelstein$^{8}$, Roman P\"oschl$^{9}$, \\ 
  J\"urgen Reuter$^{2}$, Frank Simon$^{10}$, Tomohiko Tanabe$^{1}$,
  \\ 
  James D. Wells$^{11}$ (Physics WG); Mikael Berggren$^{2}$, \\ Esteban
  Fullana$^{12}$,   Juan Fuster$^{12}$,  Frank Gaede$^{2}$,  Stefania
  Gori$^{13}$, \\
  Daniel Jeans$^{1}$,  Adri\'an Irles$^{9}$, 
  Sunghoon Jung$^{14}$,  Shin-Ichi Kawada$^2$, \\ Shigeki Matsumoto$^{7}$,
Chris Potter$^{15}$, Jan Strube$^{15,16}$\\ 
Taikan Suehara$^{17}$,  Junping Tian$^{18}$, Marcel Vos$^{12}$, \\ 
Graham Wilson$^{19}$,  Hitoshi Yamamoto$^{20}$, Ryo Yonamine$^{20}$, \\
Aleksander Filip \.Zarnecki$^{21}$ (Contributors);\\ 
  James Brau$^{15}$, Hitoshi Murayama$^{7,22,23}$ (ex officio)}

\vfill

\begin{Abstract}
To aid contributions to the Snowmass 2021 US Community Study on
physics at the International Linear Collider and other proposed $\ee$
colliders, we present a list 
of study questions that could be the basis of useful Snowmass
projects.  We accompany this with links to 
 references and resources on $\ee$ physics, and a description of a new
 software framework that we are preparing for $\ee$ studies at Snowmass.
\end{Abstract}
\vfill

\newpage
\phantom{top of page}

\vfill

\noindent { \it   \hskip - 0.05in
$^1$  \KEK \\ 
$^2$  \DESY\\
$^3$ \Berlin \\ 
$^4$ \SLAC \\ 
$^5$ \Peking\\
$^6$   \Osaka \\ 
$^7$ \IPMU \\ 
$^8$ \Cornell \\ 
$^{9}$   \Orsay \\ 
$^{10}$  \Munich \\ 
$^{11}$ \Michigan \\ 
$^{12}$  \Valencia \\ 
$^{13}$  \UCSC \\ 
$^{14}$  \SNU \\ 
$^{15}$   \Oregon\\ 
$^{16}$   \PNNL\\ 
$^{16}$   \Kyushu\\ 
$^{18}$  \Tokyo\\
$^{19}$  \Kansas\\ 
$^{20}$ \Sendai \\
$^{21}$  \Warsaw\\ 
$^{22}$  \Berkeley \\ 
$^{23}$   \LBNL }

\vfill

\newpage

\tableofcontents
\end{titlepage}

\newpage

\def\thefootnote{\fnsymbol{footnote}}
\setcounter{footnote}{0}

\section{Introduction}

Perhaps the most important question to be addressed in the 2021
Snowmass study is that of whether the US should take a major role in an
electron-positron collider as the next global project in high-energy
physics.  A primary element of the physics motivation is the
opportunity that  such a facility offers 
to explore the properties of the Higgs particle with high precision.
Electron-positron collisions also offer access to other
important physics topics, including the study of top quark and
fundamental tests of the electroweak and strong interactions.

In order for members of the US high energy physics community to grapple
with this question, it is crucial for them  to explore in some depth
how a next-generation  $\ee$ collider can advance the physics studies
that they consider most important.  Over many years,  reports and reviews have been
written about the prospects for  measurements at next-generation $\ee$
colliders.  But  there is no substitute for actually  working oneself
on a 
physics analysis, confronting the 
capabilities of an $\ee$ collider with a particular  issue in particle
physics and exploring what can be learned beyond the reach of the LHC
and other current
facilities.   The purpose of this document is to make such studies
accessible to as many members of the community as possible.   We
expect that such studies will improve our current understanding of the
best-studied measurements and also reveal new directions that can be 
explored with $\ee$ colliders. 

The possible physics measurements of  next-generation $\ee$ colliders have been studied
for the International Linear Collider (ILC) and the Compact Linear
Collider (CLIC) under the auspices of the Linear Collider
Collaboration.  These measurements have also been studied by the
Circular Electron-Positron Collider (CEPC) and the Future Circular
Collider (FCC-ee) groups.  There is a related program of detector 
R\&D in the linear collider community that is now quite advanced.
This program is  focused on detector elements needed for $\ee$ physics: high-precision
 calorimetry, extremely low-mass tracking systems, vertex detectors
 positioned close to the interaction point. These research programs
 have spun off technologies already applied in the LHC detector 
upgrades.  This report gives extensive references to these existing studies
as a starting point for future developments.   

At this moment, we see
an opportunity for the ILC to actually be constructed for operation in
the 2030's.   This makes it especially important today to understand
and evaluate the ILC capabilities.

Someone who is new to $\ee$ physics can join in studies for
next-generation $\ee$ colliders at several levels.  First, one can
study the physics analysis of $\ee$ reactions and the use of
measurements of these processes to improve our understanding of 
particle physics.   Second,  one can study in more detail the
reconstruction of particle signals in the $\ee$ environment and the
translation of the physics requirements into detailed detector
designs.  Third, one can study the current state of the art in
detector components needed for high-precision measurements at $\ee$
colliders and new technologies that could play an important role.

For most people, we expect, the easiest route into $\ee$ physics will
be to explore specific physics questions using a  fast
simulation framework or  full-simulation data in a simplified
high-level format.
 The main purpose of this document is to assist 
members of our community in joining the study at this level.  We 
will describe simulation tools and simulation data sets that
we will provide for the Snowmass study.    We then 
present a list of possible study questions that deserve attention
and that can provide a basis for thinking more deeply about the
problems of experimentation at $\ee$ colliders.  We recommend  these questions
as a way to provide  useful contributions to the physics discussions at
Snowmass 2021 and the
physics issues of next-generation $\ee$ colliders more generally.   In
many cases, the full answer can only be determined with a fully
realized detector model.  However, we expect that fast-simulation or 
simplified full-simulation studies can give insights to
estimate and minimize  specific contributions to the final
error budget. 

We also provide here resources for those who would like to study
detailed detector-related questions and questions about the underlying
technologies.  For the study of detectors, we feel it is important
make use of   the knowledge and resources of the current ILC
detector collaborations.  All of the major reactions at
next-generation $\ee$ colliders have been studied using
full-simulation tools developed by the  International Large Detector
(ILD) and Silicon Detector (SiD)  collaborations.
We recommend working with one of those collaborations and taking up an 
open project within their framework as a first step to developing new
ideas about the design and optimization of $\ee$ detectors.  This
document includes a section with the relevant 
 contact information for both detector groups.

Those people interested in the sensors and detector elements that
underlie the detector designs should recognize that these elements
are being developed by R\&D collaborations that generally involve
members of all of the  next-generation  $\ee$ collider proposals.  The
current status of the R\&D projects has recently been summarized  in an extensive
report~\cite{RDLreport}.   We recommend that people interested in
technology development should read carefully the relevant sections of
this report and reach out to the groups pursuing R\&D along the lines
of interest to them.  This document includes a section that expands
the description of $\ee$ R\&D activities  and provides contact
information for the major collaborations.

We hope that you will find this document a useful resource in
entering the community interested in research in $\ee$ collider physics.

The structure of this document  is the following:   In Section 2, we
give general references on $\ee$ physics.  In Section 3, we describe
the fast- and full-simulation frameworks that we are offering to
perform physics studies during Snowmass. In Section 4, we note some
aspects  of $\ee$ linear collider collisions that may be unfamiliar to
people who have worked only on hadron collider physics.  In Sections 5-15, we list
possible study questions related to different aspects of $\ee$
physics.  In Section 16, we give contact information for the ILC
detector concept groups.  In Section 17, we give contact information
for R\&D studies related to $\ee$ physics.

\section{General references on ILC physics}

There are many references to get started with Linear Collider
physics.  Here we highlight a few that we think are particularly
useful:
\begin{itemize}
\item 
``Primer on ILC Physics and SiD Software Tools,'' by Chris
Potter~\cite{Potter:2020kfv}
\item lectures from the Linear Collider Schools
  (\url{https://lcschool.desy.de/}), in particular, the lectures at
  the most recent schools in 2014~\cite{LCS2014}, 2016~\cite{LCS2016},
      and 2018~\cite{LCS2018}.
\end{itemize}

A comprehensive overview of the ILC physics issues and the design of
the proposed
detectors is given in the ILC Technical Design report, in particular,
in the executive summary~\cite{Behnke:2013xla} and the volumes
 devoted to Physics~\cite{Baer:2013cma}  and Detectors~\cite{Behnke:2013lya}. 

The most up-to-date detailed references on ILC physics are the papers
prepared for the European Strategy for Particle Physics 
study \cite{Bambade:2019fyw,Fujii:2019zll}.  Note that the projections
from the TDR are updated, in some cases substantially, in these
documents.    The ILD detector concept group has also produced an
updated Interim Design Report~\cite{ILD:2020qve}.

General references on the physics opportunities of other $\ee$
proposals can be found in
\cite{CEPCStudyGroup:2018rmc,CEPCStudyGroup:2018ghi} 
 for CEPC, in \cite{Abada:2019zxq},  for
FCC-ee, and in \cite{Aicheler:2012bya,Charles:2018vfv,deBlas:2018mhx},  for CLIC.
Other important recent
references are  the report of the ECFA panel on precision Higgs boson
physics commissioned as input to the 
European Strategy for Particle Physics  study~\cite{deBlas:2019rxi},
and the Briefing Book that summarizes the results of that study~\cite{Strategy:2019vxc}.

For help with ILC physics questions and the associated simulation
tools described below, 
please feel free to contact:
\begin{itemize}
\item LCC Physics Working Group conveners:  
\begin{itemize}
    \item    Keisuke Fujii
  (keisuke.fujii@kek.jp), Christophe Grojean
  (christophe.grojean@desy.de), Michael Peskin
  (mpeskin@ slac.stanford.edu)
\end{itemize}
\item ILC detector concept group physics coordinators: 
\begin{itemize}
  \item    SiD: Tim Barkow
  (timb@slac.stanford.edu)
\item ILD:  Keisuke Fujii
  (keisuke.fujii@kek.jp),  Jenny List (jenny.list@desy.de)
\end{itemize}
\item ILC contacts for the various Energy Frontier working groups
\begin{itemize}
  \item  EF01: Shin-ichi Kawada   (shin-ichi.kawada@desy.de) 
 \item EF02:  Maxim Perelstein  (m.perelstein@cornell.edu)
 \item EF03:  Roman Poeschl   (poeschl@lal.in2p3.fr)
\item EF04:  Sunghoon Jung  (sunghoonj@snu.ac.kr)
\item EF05:  Juergen Reuter  (juergen.reuter@desy.de)
\item EF08:  Mikael Berggren (mikael.berggren@desy.de)
\item EF09:  Taikan Suehara  (suehara@phys.kyushu-u.ac.jp)
\item EF10:  Aleksander Filip \.Zarnecki  (filip.zarnecki@fuw.edu.pl)
\item TF07:   Mihoko Nojiri (nojiri@post.kek.jp)
\end{itemize}
\item Technical support:  ilc-snowmass@slac.stanford.edu;  ilc-snowmass on Slack
\item ILC simulation resources for Snowmass: 
\url{http://ilcsnowmass.org} .
\end{itemize}

\section{Simulation Tools and Data Sets}

Over many years, the ILC community has developed a set of simulation
tools appropriate for full-simulation studies of $\ee$ collider
processes using detailed detector designs.  These are included in the
{\ttfamily iLCSoft} package~\cite{iLCSoft}. Its output is based on the 
persistency framework and event data model {\ttfamily
  LCIO}~\cite{Gaede:2003ip}. Any {\ttfamily LCIO} file can be read and
analyzed
 in Root after loading the corresponding shared library. Examples are
 given in  the tutorials listed below.

\begin{figure}[htb]
\begin{center}
 \includegraphics[width=0.75\hsize]{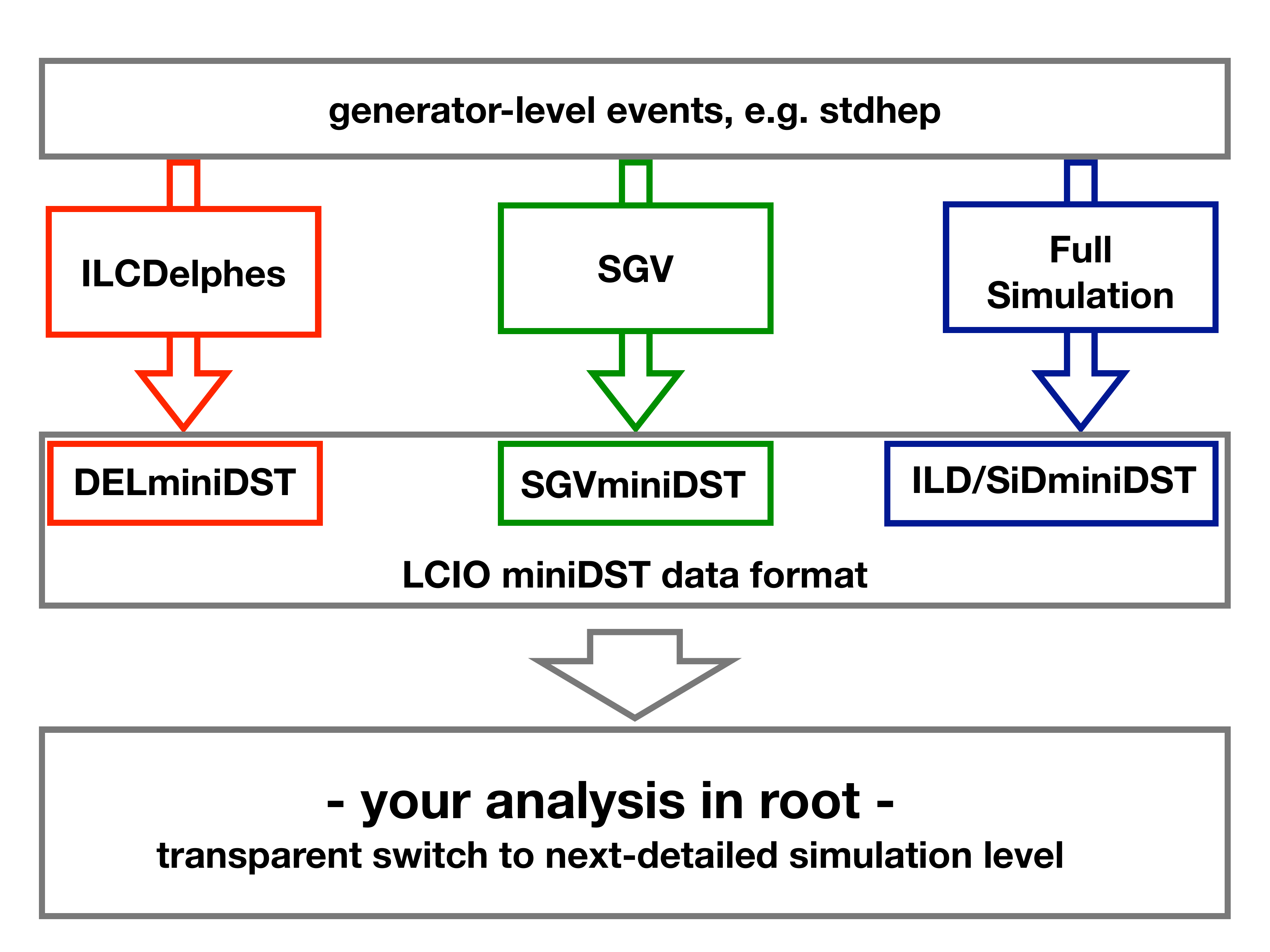} 
 \end{center}
\caption{The miniDST format can be filled via Delphes, SGV or full simulation. Analyses developed against the miniDST format can easily
switch between data from the different simulation tools.}
\label{fig:miniDST}
\end{figure}

Recently, we defined a new type of {\ttfamily LCIO} file,
the so-called ``miniDST''. As described below, the ``mini-DST'' contains
high-level reconstruction objects like jets, isolated leptons etc, but also ParticleFlow objects and MC truth information. It can be filled from three different simulation programs offering different levels of detail, as illustrated in Fig.~\ref{fig:miniDST}.

For the Snowmass study, we have developed a description of a generic
ILC detector in {\ttfamily  Delphes}~\cite{deFavereau:2013fsa}.
With the help of the Snowmass MC Task Force, we organized a series of
tutorials giving a general introduction to ILC simulation with these
tools.  The topics include:  first plots from miniDST data in
 Root and in a Jupyter notebook, the usage of the event generator {\ttfamily Whizard} 
for $e^+e^-$,  and an analysis walk-through of a search for an exotic
Higgs decay mode.  We are also making available large sets of Standard
Model events. In this section, we describe these data sets and
analysis tools. Access to the available samples, tutorials and
in-depth 
documentation is available via \url{http://ilcsnowmass.org/}. 

We hope that most members of the collider physics 
community can use these miniDST data sets, regardless whether based on {\ttfamily
    Delphes}, SGV, or full simulation, without any further introduction,
based on the provided examples. Many open questions can be explored
adequately with these formats. 
Their quality and limitations are discussed below.  On the other hand,
we will also support people who wish to dive into questions more
intimately related to the detector and the reconstruction algorithms,
which requires 
full-simulation data with
 more details than stored on the miniDST.  Again, details are given below.

\subsection{Overview on data samples and tools to be provided}
\label{tools-overview}
We are making available data samples of Standard Model events and
additional signal processes,  corresponding to a significant fraction of
the expected ILC integrated luminosity. 
The data will be provided in the following formats:

\begin{enumerate}
\item at generator-level, in stdhep format. 
   These samples can be used for generator level studies, as input
   to {\ttfamily Delphes} using the
   card describing a ``generic ILC detector'', or as input to the 
 fast detector simulation tool {\ttfamily SGV}~\cite{Berggren:2012ar}.  The
 {\ttfamily Delphes}
 and {\ttfamily SGV} tools are both described below.
\item in miniDST format.   This format, described below,  contains a 
        condensate of the high-level reconstruction output, 
   readable in Root.  We currently provide two flavors of 
minDST:   Delphes-miniDST, produced with {\ttfamily
    Delphes}, and  ILD-miniDST, produced with the ILD full simulation
  and reconstruction chain. SGV-miniDST is in preparation.
 
\end{enumerate}

Links to our software tools and data are given at \cite{ILCURL}.
The stdhep files are available on the grid vo ``ilc'' and
physically reside at DESY and KEK storage elements.  The stdhep and Delphes-miniDST files are
available from a repository at SLAC, for easy access
by the US community. If you would like to request access to
a specific data set listed below, please send your  request to  the
ILC contact persons listed in the
previous section.

Here is a summary of the  data sets and tools that we make available:

\begin{itemize}   
\item available already:
   \begin{itemize} 
   \item[DATA] generator-level event samples, stdhep format, for
     $\sqrt{s}=250$\,GeV, 
$350$\,GeV, $500$\,GeV, $1$\,TeV
  \item[DATA]  Delphes-miniDSTs for the above generator-level samples
   \item[TOOL] {\ttfamily SGV} fast simulation
  \item[TOOL] {\ttfamily Delphes} card for a ``generic ILC detector''
  \item[TOOL] {\ttfamily delphes2lcio} to run {\ttfamily Delphes} 
              with output in miniDST format.
   \end{itemize}
\item in preparation:
   \begin{itemize} 
   \item[DATA] SGV-miniDST for the above generator-level samples
    \end{itemize}
\item available on request and with ILD guest membership (see below):
   \begin{itemize} 
   \item[DATA] ILD-miniDST of new $250$\,GeV samples, other energies from previous MC production
   \end{itemize}
\end{itemize}

We request that people who use the stdhep events, the {\ttfamily
  Delphes} card, the   {\ttfamily Delphes} files or 
SGV-miniDST files, or the tool {\ttfamily SGV} cite this document along with the
specific references for individual tools. Support for the use of
 these tools can be obtained from the authors and the 
contact persons in each working group.

Studies that require detailed modelling of the detector and its
response should use full-simulation data. The full
 simulations of the SiD and ILD detector concepts and
the
 corresponding reconstruction are contained in  the {\ttfamily iLCSoft}
 package~\cite{iLCSoft}. 
Both SiD and ILD welcome
new collaborators and will support studies in the Snowmass
context.  Both ILD and SiD offer a ``guest
membership'' for institutions 
and individuals.  This has no cost 
but only requires that one follow the 
 publication rules of the concept group.  Contact information for SiD
 and ILD can be found in Section 16 of this report.

\subsection{Generator-level events}
\label{subsec:gen-level}

The existing generator-level event samples have been produced based on
 a dedicated ILC tune 
of the event generator  {\ttfamily
  Whizard\,1.95}~\cite{Kilian:2007gr,Reuter:2020inh}. They are available in stdhep
 format for center-of-mass energies
of $\sqrt{s}=250$\,GeV, $350$\,GeV, $500$\,GeV, $1$\,TeV. 
At each energy, the full Standard Model plus selected signal processes
have been 
generated.   The luminosity of a linear collider is expected to
increase proportional to $\sqrt{s}$, so we have produced event
samples corresponding to integrated luminosities increasing in the
same way,  roughly 250\,fb$^{-1}$ for 250\,GeV, up to 1000~fb$^{-1}$ for 1\,TeV.
A list of all generated processes, along with
cross sections, numbers of events, {\it etc.}, can be browsed at
\url{https://ilcsoft.desy.de/dbd/generated/}. The standard assumptions
of the ILC run plan for 
total integrated luminosity at each
energy can be found in~\cite{Barklow:2015tja}, with an update
 for $\sqrt{s}=250$\,GeV in~\cite{Fujii:2019zll,Fujii:2017vwa}.

The new event generation at 250\,GeV mentioned above will produce a
10~ab$^{-1}$ SM data sample using {\ttfamily Whizard\,2.8.3}.  This
sample incorporates a number of evolutionary improvements in
{\ttfamily Whizard} and also includes a slightly harder spectrum of
beamstrahlung radiation, as called for in the latest ILC 250 designs.

The ILC-tuned {\ttfamily Whizard} generator includes a model of the energy distribution
of the incoming beams,  giving the luminosity spectrum, for each of the
projected energies.  This was  created by the beam-beam simulation
code {\ttfamily
  GuineaPig}~\cite{Schulte:1999tx}.  {\ttfamily Whizard} includes  a full treatment of
spin information from the inital helicity of the incoming
electrons and positrons to the decay products of
$\tau$-leptons. Hadronization is performed 
by {\ttfamily Pythia6}~\cite{Sjostrand:2006za} tuned to LEP
data~\cite{BerggrenPYTHIA}. 

To study additional signal processes, for example, new particle pair
production, one can generate events using  {\ttfamily Whizard} to
include a realistic luminosity spectrum.  On the other hand, it is
usually a
reasonable first approximation to  generate events at the nominal
center of mass energy.  In either case, it is important to use the SM
backgrounds that we provide rather than generating a new set of
backgrounds.  We would be happy to help in the generation of new
signal reactions, and in the production of miniDSTs for these samples.

The data sets have been generated for pure  initial state
helicities in all four possible combinations  ($e^-_L e^+_R$, $e^-_R
e^+_L$, 
$e^-_L e^+_L$, $e^-_R
e^+_R$). A sample with any polarization  $(P_{e^-},P_{e^+})$ can be
created by drawing events from these samples with appropriate
weights. This allows one to study the effect of varying beam
polarization and to compare to the situation of perfectly polarized
beams.  The  ILC baseline is  $P_{e^-}=\pm 80\%$ and $P_{e^+}=\pm
30\%$.   We also consider an upgrade to $P_{e^+}=\pm 60\%$.  The
weights to create these polarization samples are given in
Table~\ref{tab:polweights} in Appendix A. 

\subsection{Delphes}

 We have developed a  {\ttfamily Delphes}
  card  for the Snowmass process  describing a ``generic ILC
  detector''.  We hope that this will provide an immediately
  accessible tool that mimics the response of the ILD and SiD
  detectors as closely as possible within the  {\ttfamily Delphes}
  parametrized framework.   This new description is based on, and improves upon,
earlier  {\ttfamily Delphes} cards for $\ee$
  collider physics~\cite{Potter:2016pgp,Chen:2017yel,Leogrande:2019qbe}.
 It offers information important for $\ee$ 
analyses that have not previously been available in  {\ttfamily
  Delphes}, including charm tagging and  the geometry of 
 the very forward BeamCal and LumiCal detectors.  It includes
 event interpretation with a fixed total number of jets, a standard
 method in $\ee$ physics that is
 included in the CLIC  {\ttfamily Delphes}
 distribution~\cite{Leogrande:2019qbe},
 and also widely used in ILC analyses.
  
We have also developed a tool called {\ttfamily delphes2lcio}~\cite{delphes2lcio}, which
runs {\ttfamily Delphes} and provides output in the miniDST format.
The advantage of using this output is that it will be easier to transform an analysis 
based on  {\ttfamily
  Delphes}
 to one based on a more precise description of the ILC detectors.  The new  {\ttfamily
  Delphes}
still omits some more detailed aspects of the ILC detector
description, including the presence of a beam crossing angle and the
proper correlations among flavor tags for each jet.  These effects are
included in the SGV-miniDST and ILD-miniDST
 files described below.   Any analysis code developed
against the  {\ttfamily
  Delphes} miniDST format will be easily transferrable to 
run on SGV-miniDST or  ILD-miniDST files, should 
the need for a more precise modeling of the detector 
response arise during the course of the study.

\subsection{SGV (Simulation \`a Grande Vitesse)}

The standard  fast simulation tool  used by the ILD concept group
is 
{\ttfamily SGV}~\cite{Berggren:2012ar}.  {\ttfamily SGV} is a fast-simulation
program, in the sense that the calorimeter response is para\-me\-trized,
but its description of tracking 
is derived from the specified geometry and point resolutions via a
covariance engine approach.  In practice, this means that {\ttfamily SGV} models the track
and vertex reconstruction performance, and thereby also the flavour tag performance of
the full simulation and reconstruction chain, almost perfectly. 
The calorimeter parametrization is
more detailed than that of  {\ttfamily Delphes} and is adapted to the
strategy of highly granular calorimetry and particle flow that is used
in ILC analyses.  At the same time,  {\ttfamily SGV}  is as fast as 
 {\ttfamily Delphes}  in terms of CPU time/event~\cite{BerggrenSGVperf}.

 {\ttfamily SGV} is
available for download and usage from the DESY
svn-server~\cite{getSGV}.
 It can read externally stored
events in various formats, including generator-level events in stdhep 
and in LCIO.  It can also 
 generate events on-the-fly, by calling {\ttfamily Whizard 1.95} or
 {\ttfamily Pythia6} internally. 

\subsection{miniDST}

The miniDST is a high-level data format which contains the
Monte-Carlo truth information, the reconstructed particle flow objects
and the links between truth and reconstruction level, event shape
variables, isolated electrons, muons, taus and photons, as well as the
rest of the event clustered into 
2, 3, 4, 5, and 6 jets, along with $b$- and $c$-tagging information.
For the 500\,GeV and 1\,TeV samples the clustering extends from 2 to
8 jets. By loading the LCIO shared library and including the LCIO header
file into your Root macro (or compiled program), the
miniDST can be read directly in Root. Examples of how to
 access the different types of information are provided
 here~\cite{miniDST}
 and in the Snowmass tutorial~\cite{miniDSTtutorial}. Thanks to the LCIO 
compression, the miniDST format requires about the same or even slightly less 
space per event than the stdhep format, despite the additional reconstruction-level
information.

The miniDSTs can in principle be filled from a number of sources. In
Sec.~\ref{tools-overview}, we referred to ILD-miniDSTs and SGV-miniDSTs. For
ILD-miniDSTs, the various objects are created from the full simulation
and reconstruction chain of ILD. 
 For SGV-miniDSTs, the miniDSTs are filled after simulating the detector
response with {\ttfamily SGV}. As we have explained above, this gives
a highly accurate
 representation of the full simulation,
 sufficient for most studies which can be 
done on the miniDST format. Delphes-miniDSTs are created via {\ttfamily delphes2lcio}~\cite{delphes2lcio}.

As the new, {\ttfamily Whizard\,2.8.3}-based
 $\sqrt{s}=250$\,GeV events are being processed through
the ILD full simulation, ILD produces ILD-miniDSTs from the full
simulation and reconstruction chain, depending on the demand. The existing ILD and SiD full simulation
samples for the higher energies 
could also be produced in a miniDST version, provided demand and resources.  To use  the full
simulation  miniDSTs, however, one should
get in contact  with the relevant concept group ILD or SiD.

\subsection{Matching Tools and Topics}
For each of the topics listed in the following sections, we recommend 
a {\em minimal} level of 
detail to go beyond the current state of the art.  The following
labels are used:

\begin{itemize}
\item[TH] questions of phenomenology that do not require MC simulations
\item [GEN] questions that can be studied from  generator-level events
\item [DEL] questions that can be addressed, to first order, using {\ttfamily Delphes}
\item [miniDST] questions that can be studied with miniDSTs from
 {\ttfamily SGV} or full simulation
\item [FULL] questions that require at least the full DST from full
  simulation and reconstruction, or 
detailed simulation of different detector variants, or development of 
high-level reconstruction tools.
\item [SPEC] questions that require special tools, \eg,  to simulate
  the ILC beams or to incorporate higher-order theory calculations.  
Please consult with the experts listed above.
\end{itemize}

\section{Notable features of $\ee$ collisions}

We should note two important aspects of $\ee$ physics that might be
unfamiliar to people who have worked only at hadron colliders.  The
first is that linear $\ee$ colliders will provide longitudinally polarized electron and
positron beams.   Control of the beam polarization can then be used as a
powerful tool for $\ee$
physics.  Beam  polarization has an  order-1 effect on ILC cross
sections, since the $e^-_L$ and $e^-_R$ have different
$SU(2)\times U(1)$ quantum numbers.  The ILC expects to provide 80\%
polarization  in the electron
beam and 30\% polarization in the positron beam, with the possibility
in both cases of rapidly switching the polarization orientation.  This
effectively quadruples the number of observables.  Some of the
polarization asymmetries provide new physics information; others allow
cross-checks for the estimation of systematic
errors~\cite{Fujii:2019zll,Fujii:2018mli}.

The second is that the nominal center of mass energy of $\ee$
collisions is affected both by initial-state radiation and by
radiation from the beam-beam interaction (``beamstrahlung'').
Beamstrahlung and ISR have three important effects.  First, they
broaden the $\ee$ center of mass energy distribution.  This broadening
is a few percent at energies up to 500~GeV.  This effect   is included
in all of the  samples that 
 we provide;  see
Sec.~\ref{subsec:gen-level}. More importantly, ISR and beamstrahlung
produce photons that  induce hard $\gamma\gamma$ and $e\gamma$
reactions.  Those processes are often the major source of  background
events, in  particular for many types of searches.
They  are included in the full SM event samples  we
provide, and they
     should be included by default, along with $e^+e^-$-induced
     backgrounds, in all physics analyses. 
Finally, ISR and beamstrahlung radiation contribute an additional ``overlay''
background due to independent low-$\sqrt{s}$ electron-photon and 
photon-photon reactions, similar to pileup at hadron colliders.
The rate of these events is energy dependent. At a center-of-mass 
energy of $500$\,GeV, for instance, they lead to, on average, $1.1$ 
events of overlay background.   This background source---mainly, very
soft hadrons and leptons---has been shown to be well-manageable 
even at a center-of-mass energy of $1$\,TeV~\cite{Behnke:2013lya}, and
plays hardly any role at $250$\,GeV. 
Its residual effect will be included in the ILD-miniDSTs.

Finally, we emphasize the interest in considering all of the topics below within
the context of the energy evolution of ILC.   The initial stage of ILC
will be at a CM energy of 250~GeV.   Once the ILC infrastructure is in
place and significant physics results have been obtained, it will be
almost imperative to upgrade the machine to a CM energy of 500 or 550~GeV,
opening up the precision study of the top quark, measurement of
the top quark Yukawa coupling, measurement of the Higgs boson
self-coupling, and an increased search region for new particles.
Official ILC parameter sets have been presented for CM energies of
91~GeV, 250~GeV, 350~GeV, 500~GeV, and 1~TeV~\cite{Fujii:2019zll}.
Especially at the lowest and highest of these energies, there are many
issues that are still unexplored.

\section{Questions about general $\ee$ event analysis}

\begin{enumerate}
 \item QCD generators. So far, most LC studies have been
  performed with signal and backgrounds generated by 
leading-order parton-level generators (especially, Whizard),
followed by parton showering and hadronization in Pythia6 tuned to LEP
data.   Details of the  tune used in studies for ILD are
described in \cite{BerggrenPYTHIA}.
Is this  approach adequate for the high-precision jet measurements
needed for ILC physics studies and, if not,
    what should replace it? What additional experimental measurements
    would be helpful as inputs to the tuning?  [GEN] 

\item Jet reconstruction. In the past few years, there have been 
significant improvements in the
  theory of
 jet reconstruction, and current ILC studies do not yet take advantage of
 these.  Jet reconstruction in $\ee$ collisions is
 different from  that at 
 hadron colliders.  At $\ee$, we have good knowledge of the CM energy
 and
 forward-backward momentum
 balance, so typically we analyze jets with full 3-dimensional
 information and assign the whole energy-momentum measured by the
 detector to jets.  This affects both the distance criteria and the
 clustering schemes themselves.  On the other hand, current jet reconstruction
 algorithms used in $\ee$ studies~\cite{Suehara:2015ura} are still similar to those from 
 the LEP era.  At the same time, calorimetry is expected to be much
 improved at future $\ee$ colliders, so that the 2-jet invariant mass resolution will be
 dominated not by detector resolution but rather by
 mis-clustering~\cite{Boronat:2016tgd,TianJet:2018}.   Can we use our
 new understanding of jet structure in QCD to develop new  clustering algorithms
 with higher fidelity?    [GEN]

\item Tau identification and reconstruction.  Taus can be
  reconstructed at $\ee$ colliders by identifying their  individual decay
  products without relying on a jet algorithm.  Still, high fidelity
  is needed, so it is important to consider how to optimize the
reconstruction of tau events, both to maximize efficiency and to minimize systematic
  errors.  Current reconstruction algorithms are described in
  \cite{Kawada:2015wea,Jeans:2015vaa}.  The second of these extracts
  the 
tau neutrino momentum using kinematic constraints enabled by
  excellent
 resolution on impact parameters.  [miniDST]

\item Tau polarization.  How accurately can one measure tau
  polarization, both in Higgs decay ($p_{lab} \sim 50$~GeV) and at
  500~GeV ($p_{lab}\sim 250$~GeV)?  What are the dominant systematic
errors?  How can these be minimized?   Current studies can be found in
\cite{Jeans:2018anq}, for Higgs decays, and in \cite{Jeans:2019brt},
for $\ee\to \tau^+\tau^-$ at 500~GeV.    [FULL]

\item  Lepton reconstruction.  What detector specifications are needed to
  capture and recombine final-state radiation from leptons (including
  taus), and what 
are the resulting
  systematic errors on the lepton energy?  An algorithm for 
recovery of bremsstrahlung and final state radiation in $Z\to
\ee$ and $\mu^+\mu^-$ events is included in the 
 ILD reconstruction~\cite{Wendt:2007iw}.  [FULL]

\item Luminosity measurement from Bhabha scattering.  At the ILC, it
  is necessary to measure with high precision both the
  luminosity and the luminosity spectrum, which is a function of the
 actual  $\sqrt{s}$ of the $\ee$ annihilation.  It is envisioned that this will be
  done primarily by measuring small-angle Bhabha scattering.  Current
  studies 
project uncertainties of a few per mille on the luminosity measurement
at 500~GeV and at 1~TeV, where the leading contribution originates from
the modeling of the 
electromagnetic deflection of the Bhabhas in the electric field of the 
outgoing bunch~\cite{Bozovic-Jelisavcic:2013aca}.    The systematic
uncertainty 
on the shape of the luminosity spectrum has been 
studied for CLIC in~\cite{Poss:2013oea} and for ILC
in~\cite{Sailer:2009zz}. 
  What uncertainty should be
expected at 250~GeV and 350~GeV, in the absolute luminosity and
the shape?  Can the measurement strategy be improved?   [SPEC]

\item Luminosity measurement from other processes.     The reaction
  $\ee\to \gamma\gamma$ 
is also sensitive to the luminosity spectrum and could also
contribute to that spectrum measurement.   What improvements are
achievable from the use of this reaction?   Are there other reactions
that could contribute significantly?   [GEN]

\item  Implications of luminosity spectrum uncertainty.   In questions
  below, we ask about the effect of the luminosity spectrum on
  measurements of the Higgs boson mass and the top quark mass.  Are
  there other measurements that are affected in an important way by
  the uncertainty in the  luminosity spectrum?   What uncertainties  on the
  luminosity spectrum are required to bring these sources of error
  under control?   [GEN]  

\item Precision calorimetry.   Early studies of the physics case for
  $\ee$ colliders emphasized precision hadron calorimetry ($\sim
  25\%/\sqrt{E}$)  to be able to separate $W$ and $Z$ bosons in their
  hadronic decay modes.  This is an important motivation at 500~GeV and above, but
  less so for  the initial 250~GeV program focusing on $\ee\to Zh$.
  What performance is required for hadron calorimetry at to meet the
  needs of the physis at 250~GeV?  A 
 recent discussion of this issue is found in \cite{ManqiCal}.  [Delphes]

\item Photon energy resolution.   The SiD and ILD detector designs
  achieve excellent  jet energy
resolution  by using
       highly granular electromagnetic sampling calorimeters. This comes at the
expense of
       intrinsic photon energy resolution.   Is there a better
  optimum with improved photon energy resolution or photon-finding
  efficiency?  What are the advantages and disadvantages?    [FULL]

\item  Heavy Flavor tagging and vertex charge measurement.  The SiD and ILD
detectors are designed to have excellent efficiency for  secondary
vertices.  The currently expected heavy
  flavor-tagging performance is described in \cite{Suehara:2015ura}. 
It is also expected that these detectors can reconstruct the vertex
charge, to discriminate heavy quarks from
antiquarks~\cite{Bilokin:2017lco,BilokinTBA}.  Can the performance on
these quantities be improved by further detector optimization or more 
 advanced machine learning
techniques?     [FULL]

\item   Particle ID.    The SiD and ILD detectors do not have dedicated
  subdetectors for particle ID; however, they can identify kaons by
  $dE/dx$ measurement or by track timing.  Identification of kaons is 
important for many aspects of $\ee$ physics, including the vertex
charge measurement described in the previous question 
 and the reconstruction of $\tau$ leptons from their
final states.  It is then interesting to consider  how kaon
identification could be improved, for example,  using the new timing detector
technologies being
developed for HL-LHC, and the
implications for $\ee$ physics measurements.  [miniDST]. 

\item  Strange quark  tagging.  It would be useful to be able to
  tag strange quark jets in $\ee$ processes.
  Studies  of $Z\to s\bar s$ from the LEP/SLC era can be found in
  in~\cite{Stangle,Abreu:1999cj}; these use dedicated RICH/CRID
  particle ID detectors.   More recent proposals for  strange
  taggers, for general purpose detectors, are 
  given in~\cite{Kats:2016ghi,Duarte-Campderros:2018ouv,Nakai:2020kuu}. 
These strategies can surely be improved.  [miniDST]

\item  Tracking detector momentum scale.
How well can a modern detector such as SiD and ILD be designed so as
to control the momentum scale at levels which are interesting for
a precision center-of-mass energy determination?
This involves issues
of magnetic field-mapping, alignment and tracker design. Can one
approach the limitation set by the knowledge of the $J/\psi$
at 2 ppm? Are there other particles that can be used as additional
momentum standards?  Some motivations for this study are 
described in \cite{Fujii:2019zll}.    [miniDST]

\end{enumerate}

\section{Questions about Higgs boson physics: $\ee \to Zh$}

\begin{enumerate}

\item Total cross section.  The total cross section for $\ee\to Zh$
 can be measured cleanly using
  leptonic $Z$ decays, but with much higher statistics using hadronic
  $Z$ decays. For the ILC at 250~GeV, this cross section is expected
  to be
  measured to 0.8\%.   It is a key input to all Higgs coupling
  measurements. 
 How can the Higgs events best be separated from $\ee\to
  ZZ$ and other backgrounds?  Current studies can be
  found in \cite{Yan:2016xyx} for the  leptonic recoil channel,
    and in \cite{ Miyamoto:2013zva,Tomita:2015, Thomson:2015jda},  for
    the hadronic recoil channel.   [miniDST]

\item Model-independence of the total cross section measurement.  It
  is important that the total cross section measurement be independent
  of the Higgs decay final state to the greatest extent possible.
  This issue is studied for SM Higgs decays in
  \cite{Yan:2016xyx,Thomson:2015jda}.  Can this model-independence be
  extended to the various possible types of exotic Higgs decays?    [DEL]

\item Higgs boson mass -- leading channel.  It is important to measure the Higgs boson
  mass with as high a precision as possible.  For ILC at 250 GeV, the
  expected statistical uncertainty is 14~MeV, using the recoil $Z$
  decay $Z\to \mu^+\mu^-$~\cite{Yan:2016xyx}.   List and estimate the
  sources of systematic error in this measurement.
  Sources to consider include  the luminosity spectrum uncertainty, the
  tracker momentum scale, and the subtraction of $\ee\to ZZ$ and other
  backgrounds.  Is there a method for in-situ energy calibration?   Our
  current estimate of the systematic error is 20-30~MeV, from the
  comparison to the kinematics of $\ee\to ZZ$. 
  Is there a complementary way to obtain this information?     [miniDST]

\item Higgs boson mass -- other channels.  Is it possible to make
  competitive measurements of the Higgs boson mass using other $Z$
  decay channels?  In particular, improvements in electron
  reconstruction might make $Z\to \ee$ more competitive.  For the
  current status, see \cite{Yan:2016xyx}.   [FULL]

\item Higgs boson mass -- continuum.   Is it possible to make a
  competitive measurement of the Higgs boson mass by  reconstruction
  of the decay products in modes such as $h\to b\bar b$?  This
  question could be investigated both for $\ee\to Zh$, calibrated by
  $\ee\to ZZ$, or for $\ee\to \nu\bar \nu h$, calibrated by single $Z$
  and $W$ production.  A current study using $h\to b\bar b$ is given
  in \cite{Junpinghbb}.  [miniDST]

\item Invisible width.  To maximize the event sample, invisible decays
  of the Higgs should be measured with the recoil $Z$ decaying
  hadronically.  How can one best suppress the backgrounds?  What
  systematic errors result?   Current studies are given in
  \cite{Ishikawa:2014,Kato:2020pyl}.  [miniDST]

\item Higgs decays to 2 jets. At $\ee$ colliders, Higgs decays to all
  hadronic modes can be observed directly. Current studies of $h\to
  b\bar b, gg, c\bar c$ (\eg, \cite{Ono:2013sea})  date from the era before deep learning, and
  before the understanding of $q/g$ jet separation gained from LHC.
  What, now, is the optimum method for separating these three decay
  modes.  What systematic errors can be achieved?  [miniDST]

\item  Higgs decays to light quarks.  One can add to the previous
  question the possibility of Higgs decay to $s\bar s$, $d\bar d$,
  $u\bar u$.  What limits on these modes can be achieved?  Can $h\to
  s\bar s$, with $ BR = 10^{-4}$ in the SM, be observed?  A
  theoretical study on the discrimination of light quark and gluon
  initiated jets can be found  in \cite{Gao:2016jcm}. The possibility
  of observing $h\to s\bar s$ is discussed in
  \cite{Duarte-Campderros:2018ouv}
 and in question
  \#12 of Sec. 5.  [miniDST]

\item Higgs decay to $WW^*$.  In the SM, $h\to WW^*$ proceeds
  through the vertex $hW_\mu W^\mu$, but in BSM models, the
  alternative structures involving the $W$ field strength $h
  W_{\mu\nu} W^{\mu\nu}$ and  $h\eps_{\mu\nu\lambda \sigma} W^{\mu\nu}
  W^{\lambda\sigma}$ can also appear.  What is the optimal strategy
  for measuring separately the strength of these three operators?
  Which is most useful, the purely leptonic, semi-leptonic, or fully
  hadronic $W$ decay modes?   This question and the following one have
  recently been studied in the Ph.D. thesis~\cite{Ogawa}.   [DEL]

\item Higgs coupling to $ZZ^*$.  The $hZZ$ coupling can have three
  components similar to those just described for the $hWW$ coupling.
These can be measured from the reaction $\ee\to hZ$, including
forward-backward and polarization asymmetries.  How can the full
information be combined optimally?    [DEL]

\item Higgs decay to $\tau^+\tau^-$.  How accurately can one measure
  the polarization correlation in $h\to \tau^+\tau^-$ decay?  
  Studies of the sensitivity to CP-violating $h\to \tau^+\tau^-$ decays and be
  found in \cite{Jeans:2018anq,Harnik:2013aja,Berge:2013jra}.   Can the
  results be improved by making use of  better modelling of  $\tau$
  decays, using the large samples from Belle II and from ILC itself?   [miniDST]

\item Higgs coupling to $Z\gamma$. The measurement of the $hZ\gamma$
  coupling is useful to remove correlations in a global effective
  field theory fit  to Higgs
  boson parameters.  HL-LHC will give us a measurement of the $h\to
  Z\gamma$ signal strength, but another source of information might be
  the cross section for $\ee\to \gamma h$.  Is it possible to observe this
  process at the ILC  at 250~GeV?   What measurement accuracies or limits
  might be expected?   Current studies are given in
  \cite{Aoki:2019sht,Aoki:2020tba}.  [DEL]

\item Flavor-violating Higgs decays.  What limits can be placed on
  $h\to \tau \mu$, $h\to b s$, and other flavor-violating fermion
  combinations?      [miniDST]  

\item Exotic Higgs decays.   A general study of the visibility of possible exotic Higgs
  decays at $\ee$ colliders  given in~\cite{Liu:2016zki}
  surveyed  possible exotic Higgs decays and estimated the limits that can
  be achieved at $\ee$ Higgs factories.  Do a deep dive into one of
  the modes considered  (\eg, $h\to b\bar b$ + invisible light scalar).  What
  are the non-Higgs backgrounds?  To what extent is this mode
  separable from the related SM model (\eg, $h\to b\bar b$)?  Would
  this mode be discovered as an enhancement of that mode or as a
  distinctly different final state?  What limit, and what discovery
  $BR$, might be expected?     [DEL] 

\item Higgs decays to long-lived particles.  A type of exotic Higgs
  decay that deserves special attention is the possiblity of decay to
  long-lived particles.  A survey of the possibilities and discussion
  of searches for displaced-vertex signatures can be found in
  \cite{Alipour-Fard:2018lsf}.   Further studies that are more
  specific about detector requirements and elimination of
  machine-related  backgrounds would be very useful.      [FULL]

\end{enumerate}

\section{Questions about Higgs boson physics: $WW$ fusion
 and higher energy
  reactions}

\begin{enumerate}

\item Total cross section.  How can one best measure the total cross
  section for the $WW$ fusion reaction $\ee\to \nu\bar\nu h$, making use of branching ratios
  from ILC data at 250~GeV?  If one observes only the Higgs decay
  products, what is the appropriate fiducial region to give the best
  cross section estimate?  A current study can be found in
  \cite{Durig:2014lfa}.  [miniDST]

\item Higgs coupling to $WW^*$.  Can one measure the separate
  components of the $hWW$ coupling using $WW$ fusion?  What are the
  sources of systematic error?  Are there specific tests for the
  presence of the CP-violating structure?   This question is related
  to question \#9 of the previous section and is also studied in 
 \cite{Ogawa}.    [DEL]

\item $ZZ$ fusion.   The reaction $\ee\to \ee h$ ($ZZ$ fusion) has
  a much smaller cross section than  $WW$
  fusion.  However, this process may be  independently interesting as a
  probe of the $hZZ$ coupling, as described in \cite{Han:2013kya}.
  What analysis strategy allows us to make the best use of it?   This
  process can also give insight into the structure of the $hZZ$
  coupling. In that aspect, it is  related
  to question \#10 of the previous section and is studied in 
 \cite{Ogawa}.    [DEL]

\item Higgs self-coupling at 500 GeV. At 500 GeV, the Higgs
  self-coupling can be extracted  from double 
Higgsstrahlung, $\ee\to Zhh$.  This determination would be 
 model-independent, as discussed in \cite{Barklow:2017awn}.
 The current ILC projection for the expected 4~ab$^{-1}$ sample  is a 27\%
measurement of the self-coupling, limited by the efficiency of the
selection of signal from background. The
most complete current analysis, described in the Ph.D. 
thesis~\cite{Duerig},  identified many parts of the
event reconstruction and selection where significant improvements are
possible.   These include  the flavour tag, the jet
clustering, the identification and correction for semi-leptonic decays in
$b$ jets, the use of squared matrix elements  as selection
variables, and 
the inclusion of $h/Z \to \tau^+\tau^-$ channels.  Many of these
elements could benefit from the use of advanced machine-learning
techniques.  How much improvement is possible?     [miniDST]

\item Higgs self-coupling at higher energies.  The same questions can
  be asked about the process $\ee\to \nu\bar\nu hh$, which turns on at
  higher energy.  This is projected to yield a 10\% measurement of the
  Higgs self-coupling with the expected  8~ab$^{-1}$ event sample for
  ILC at   1~TeV~\cite{Tian:2013qmi}.  Studies for CLIC project a 9\%
  accuracy from measurements at 1.5~TeV and
  3~TeV~\cite{Roloff:2019crr}.  Again, the limiting factor is
  signal/background discrimination.  How can this be 
improved with more sophisticated tools?   What is the
  lowest CM energy at which such a precise measurement can be
  achieved?      [miniDST]

\item Off-shell Higgs couplings.  Most ILC reactions involve the Higgs
  boson on mass shell.  Are there reactions that can test the Higgs
  boson couplings at off-shell momenta, which can give additional
  information to test some BSM models?   [GEN]

\end{enumerate}

\section{Questions about top quark physics}

A recent overview of top quark physics at $\ee$ colliders can be found
in~\cite{Vos:2016til}.

\begin{enumerate}

\item Top quark threshold. The top quark pair-production
 threshold is a sudden jump in the $\ee$
cross section over an interval of a few GeV. Its detailed shape
depends on $m_t$, $\Gamma_t$, $\alpha_s$, and the top quark Yukawa
coupling $y_t$. It is washed out, to
a certain extent, by initial-state radiation, beamstrahlung, and machine energy
spread.  Because the threshold is sharp, we expect a measurement of
the top quark mass with a statistical uncertainty better than 20~MeV.
The dominant error comes from the theory of the threshold shape.
 The extraction of the physics parameters and the optimal
scanning strategy has been studied for some time;  see
\cite{Comas:1995rw,Simon:2016pwp,Simon:2019axh,Nowak:2019tjy} 
and references therein. Some issues that still need further study are:
Can the precision in these parameters be improved by 
using the dependence of the cross section on beam polarization? 
Can the precision be improved using properties of the observed final
state $b\bar b W^+W^-$?  [GEN]

\item Top quark threshold -- unfolding:  Effects of initial state- and beamstrahlung
  radiation and machine energy spread on the top quark threshold shape
  deserve further study.    To what
extent can one use measurements of these effects to better unfold the 
underlying threshold shape?   [SPEC]

\item Top quark mass from pair production.  It is also possible to
  obtain a precise top quark mass from measurements of
 the kinematics of top quark pair-production.   There is an
 interesting theoretical literature on this question
 \cite{Hoang:2017kmk, Dehnadi:2018hrh}, which explains that a
 short-distance top quark mass such as the $\MSB$ mass  can be
 obtained from measurements well above threshold.  However,
  little attention has been given to the experimental aspects of this
  program.  A method for determining the top quark mass from $\ee\to
  t\bar t \gamma$ events has been  studied in \cite{Boronat:2019cgt}.
  What is the best strategy to extract a
  short-distance top quark mass with high precision from pair-production events?
 [miniDST]

\item  Top quark spin.  The spin of the top is transferred to its decay
  products before hadronization and can thus be measured from the
  final state of top decay.  How well can one measure the top quark
  spin and the $t\bar t$ spin correlation, for example, as a function
  of the production angle?  How does the accuracy of this measurement
  compare to that at hadron colliders? 
[miniDST]

\item Top quark form factors.  There are 4 possible form factors for the
  top quark coupling to each of $\gamma$ and $Z$, with one, in each
  case, CP-violating.  How can one best make use of beam polarization, the
  choice(s) of 
 the center of mass energy, and the measurement of $t$ and $\bar t$
 spins to determine these form factors independently?
  What is the most powerful signal of top quark-associated CP
  violation at the ILC?  Current studies of these questions are given
  in \cite{Amjad:2015mma,Khiem:2015ofa,Bernreuther:2017cyi}. A more
  expansive version of this question, in the context of SM Effective
  Field Theory, is given in question \#6 of Sec. 15.  [GEN]

\item Form factors near threshold.   At threshold, the combinations of
  top quark form  factors that appear in
  the spin-1 S-wave are dominant, while orthogonal combinations are
  suppressed.  To what extent can the separate form factors be
  measured?   How can beam polarization assist in measuring these form
  factors?  How can the CM energy-dependence, as one moves through the
  threshold, be used?   [TH]

\item Top quark width.  How well can the top quark width be measured
  directly at the ILC?  How would we best combine the determinations
  from the threshold region and from higher energies?   [GEN]

\item Higgs coupling to $t\bar t$. The top quark Yukawa coupling is very important to
measure precisely. The measurement of this quantity at 500~GeV is limited
by the fact that this energy is very close to the threshold for the reaction
$\ee\to t\bar t h$. The precision is expected to improve to about 2\% at
550--600~GeV,
 from simple cross section scaling~\cite{Barklow:2015tja}.
Can this be confirmed by a full study?    [DEL]

\end{enumerate}

\section{Questions about $\ee\to f\bar f$}

\begin{enumerate}

\item  New vector resonances.  There is an extensive literature on
  searching for signals of new $s$-channel resonances through
  deviations of the $\ee\to f\bar f$ cross sections from the SM
  expectations~\cite{Zpreview}.  However, the theoretical landscape changes with time,
  and much phase space available to earlier studies has now been
  excluded by measurements at the LHC.  What is the current situation?  [TH]

\item New observed vector resonances.   There is substantial phase
  space for the discovery of a BSM resonance in Drell-Yan production
  at the HL-LHC.  Imagine the discovery of a resonance at a mass of
  4~TeV.  What new information would the measurement of $\ee\to f\bar
  f$ at the various ILC stages bring?  How close can we come to
  completely characterizing this resonance using all available
  information from HL-LHC and ILC?   [TH]

\item $\ee\to b\bar b$.   In BSM models in which the top quark plays
  a role in the dynamics of a composite Higgs boson, there must also
  be BSM effects on the $b$ quark, and these  might be visible in $\ee\to
  b\bar b$ at high energy. A current study at 250~GeV 
can be found in~\cite{Bilokin:2017lco}. A related study of $\ee\to
c\bar c$ is given in \cite{Irles:2020gjh}.  What information can we obtain from this
  reaction that would complement the ILC studies of the top quark?    [TH]

\item   $\ee\to \tau^+\tau^-$.   Just as for the top quark, the
  polarization of $\tau$ leptons in $\ee\to \tau^+\tau^-$ can be
  measured from the $\tau$ decay final states.  
 A current study at 500~GeV can be found
 in \cite{Jeans:2019brt}. How can we best use this additional handle to
  constrain or discover BSM models?    [TH]

\end{enumerate}

\section{Questions about $W$ boson physics}

\begin{enumerate}

\item $W$ boson mass.  It is possible to improve the precision of the
  $W$ boson mass by studies of $W^+W^-$ and single $W$ production at
  250~GeV.  Some strategies are explored in \cite{Fujii:2019zll}.  How
  can we further improve the systematic uncertainties from these and
  other possible techniques?     [miniDST]

\item  Complete event reconstruction and analysis for $\ee\to W^+W^-$.   There are 3
  possible form  factors for the
 $W$ coupling to each of $\gamma$ and $Z$, with 1, in each
  case, CP-violating~\cite{Hagiwara:1986vm}.   An $\ee\to W^+W^-$ event
  can be completely reconstructed, with all 5 decay angles determined
  in each event (up to a front-back ambiguity in the case of hadronic
  decays).  In the older literature, it is shown that even a set of 14
  complex form factors can be disentangled using optimal
  observables~\cite{Diehl:2002nj}.  However, current full-simulation
 studies~\cite{Barklow:2001ep,Marchesini:2011aka,Rosca:2016hcq} use
 only 3 angles.  It would be interesting to revisit this question and
 understand how much detailed information is available in practical
 measurements.    [DEL]

\item   Beam polarization and $\ee\to W^+W^-$.  The SM cross sections
  for $\ee\to W^+W^-$ using $e^-_L$ and $e^-_R$ beams have a
  completely different form both in the production angle and the final
  $W$ polarizations.  This implies that there must be a strong
  advantage to using polarized beams for measurements of the 
 the $W$ form factors.  Can you quantify this?
 Are there measurements that require beam polarization?   [GEN]
 
\item  CP violation in $\ee\to W^+W^-$. What are the strongest signatures of CP
  violation in the $W$ system?  What are the key observables linear in
  CP-violating parameters?   [TH]

\item $W$ polarization from hadronic decays.  There is an ambiguity in
  the determination of the $W$ decay angle between the quark and
  antiquark directions  when a $W$ decays to
  light quarks.  However, when a $W^+$ decays to $c\bar s$, the quark
  direction can be determined by charm tagging. 
   Even for $W^+\to u\bar
  d$, measurement of the final jets distinguishes transverse from
  longitudinal $W$ bosons.  What is the quantitative effect of the use
  of the quark directions on the 
  measurement of $W$ form factors?  How can this capability be used in
  other aspects of $W$ physics?      [TH]

\item Interaction of $W$ couplings and precision electroweak.  Almost
  all current studies of $W$ form factors --- both for $\ee$ and $pp$
  colliders -- assume that the couplings of SM gauge bosons to
  fermions take their SM values.  However, this approximation might
  not be warranted as the precision on the $W$ form factors improves.
 Deviations of the $e\nu W$ and $eeZ$ couplings can lead to effects
 on the $\ee\to W^+W^-$ cross section that grow as $s/m_W^2$.  How
 well do we need to know these electroweak couplings to meet the goals
 of future $W$ boson measurements?    [TH]

\item LHC constraints.  Non-standard couplings of the $W$ are already
  significantly constrained by measurements at the LHC.  How should
  these constraints be compared to the ILC capabilities?  What
  model-dependence present in the LHC limits can be removed by ILC
  measurements using events with  fully reconstructed kinematics? How does beam
  polarization enter this comparison?    [TH]

\end{enumerate}

\section{Questions about precision electroweak measurements}

Precision electroweak observables can be addressed at the ILC both
through measurements at high energies and through a dedicated run at
the $Z$ resonance.  For details, see~\cite{Fujii:2019zll}.

\begin{enumerate}

\item Radiative return.   Even before the ILC carries out a dedicated
  program of measurements at the $Z$ resonance, it can improve the
  current determination of $\sstw$ by measuring the polarization
  asymmetry of the radiative return reaction $\ee\to Z\gamma$.  The
  expected 
  statistical precision on $\sstw$ is of the order of
  $10^{-4}$~\cite{Fujii:2019zll,Ueno}. 
 What
  are the sources of systematic error?  Estimate these, taking into
  account that most events are measured by the detectors in the
  forward region.   [miniDST]

\item  ILC ``Giga-Z'' program.   The ILC could carry out a program 
  of measurements on the $Z$ resonance, collecting $5\times 10^9$
  polarized $Z$ bosons~\cite{Fujii:2019zll}.   This sample would be
  200$\times$ larger than that from LEP and $10,000$ times larger than
  that from the polarized $Z$ program at SLC.  It is interesting to
  assess this program in relation to the program proposed for circular
  $\ee$ colliders, with a much larger event sample but no beam
  polarization.  In particular, how close do the measurements in each program come to
  the expected level of systematic errors?  [DEL]

\item  Forward-backward asymmetries.   In the ILC program on the $Z$
  resonance, it is possible to
  obtain powerful constraints on the $Zq\bar q$ couplings by measuring
  forward-backward asymmetries with polarized beams.  The quark and
  antiquark directions would be determined by jet charge, or, for
  heavy quarks, vertex charge (see question \#10 of Section 5).
  However, one must assume that 
  the jet directions measured in 2-jet events (in an
  appropriate fiducial region)
  are aligned with the initial quark directions, which ceases to be
  true at higher orders in QCD.   What is the systematic
error on the measurement of forward-backward asymmetries
 due to QCD uncertainties in jet
formation and hadronization?   [GEN]

\item Tau polarization.  The determination of $\sstw$ from tau
  polarization depends on unambiguously identifying the tau decay
  mode, since each mode has its own characteristic dependence on
  polarization~\cite{ALEPH:2005ab}.  In a detector of the quality of foreseen ILC
  detectors, how well can the various tau decay modes be separated?  A
  current study is given in \cite{Jeans:2019brt}. 
What is the implication for the systematic error on the $\sstw$
  measurement obtained from $Z\to \tau^+\tau^-$?    [miniDST]

\item Flavor at the $Z$ pole.    The dedicated ILC run at the $Z$
  resonance is expected to produce $5\times 10^9$ polarized $Z$
  bosons.  This must offer unique opportunities for heavy flavor
  physics, but there have been few studies of these possibilities.
  The physics opportunities of the sample of $10^8$ polarized
  $\Lambda_b$ baryons are explored in \cite{Hiller:2001zj}.    [TH]

\end{enumerate}

\section{Questions about QCD and jets}

\begin{enumerate}

\item Jet shapes and jet substructure.   There is now an extensive
  literature on the shapes and substructure of QCD jets motivated by
  studies of jets at the LHC~\cite{Larkoski:2017jix,Asquith:2018igt}.   
This theory can be tested much more
  stringently at ILC, using the known CM energy, the absence of
  underlying events and pile-up, and the higher precision
  calorimetry.  What level of precision is possible here?  What level
  of precision can be achieved in the measurement of $\alpha_s$? 
  Can we study effects of the $b$ and $c$ masses?  Are
  there interesting BSM models that can  become visible through these
  measurements?  [GEN]

\item  Hadronization.   The large sample of 2-jet events that the ILC
  will make available offers the opportunity to test and improve
  models of hadronization.  What can be learned beyond the knowledge
  that we gained from LEP?   Specific physics topics that need new
  data are:  flavor production in jets, and characterization of $s$-
  and $g$-initiated jets; baryon production;  polarization of vector mesons (especially,
  $D^*$) and baryons in jets.   To what extent can this improved
  information feed back into improvements in LHC event analysis?    [TH]

\item Tests of parton showers.   Simulations of parton showers now aim
  for NLO and even NNLO accuracy
 (\eg, ~\cite{Hoeche:2015vea,Hoche:2017hno,Monni:2019whf}).  How well
 can we test the accuracy of parton shower generators at $\ee$
 colliders, both for their general accuracy in reproducing event
 shapes and for specific modelling of features of QCD that appear at
 high order?  [GEN]

\item Structure of gluon jets.  The Higgs production processes in
  $\ee$ with
  the decay $h\to gg$ gives a clean, low-background sample of
  gluon-initiated jets.
 A study  of the QCD structure of this final state can be
found in \cite{Gao:2019mlt}.  
  How can we use this sample to improve our knowledge of gluon jet
  substructure and nonperturbative gluon fragmentation?    [GEN]

\item Structure of top quark final states.  The reaction $\ee\to t\bar
  t$ gives a well-characterized and almost background-free sample of
  top quark events.  How can this be used to improve our knowledge of
  QCD jet structure?   [GEN]

\end{enumerate}

\section{Questions about searches for new particles}

\begin{enumerate}

\item  Light Higgsino.  In the MSSM, the Higgsino can be light
  compared to the other superpartners, and the splitting between
  Higgsino mass eigenstates is then naturally less than 5~GeV.  In
  this parameter region, it is very difficult to discover the Higgsino
  at the LHC, while the ILC can be a Higgsino factory.  Some studies
  of Higgsinos at the ILC can be found in
  \cite{Berggren:2013vfa,Baer:2014yta,Baer:2019gvu,PardodeVera:2020zlr}. 
 The signatures change
  rapidly  as a function of the $\widetilde{h}^+$-$\widetilde{h}^0$
  mass splitting, so it is interesting to extend these studies and
  explore the full range of this parameter.    [miniDST]

\item Light or compressed SUSY scenarios:  More generally, once the
  masses of   colored superpartners are taken to be above  the reach of the LHC, the signatures
  of color-neutral superpartners and the corresponding search
  strategies depend on the fine details of the SUSY parameter set.
  Models with large mass gaps between the wino ($\s w$) and bino ($\s b$) and models
  with sleptons below the wino mass are relatively accessible at the
  LHC, but, other choices are more difficult.  This issue has been analyzed
 in \cite{Berggren:2020tle}.   Pick up one of the
  scenarios discussed in this paper (for example, models with a small
  lepton-neutralino mass difference),
 and make a detailed comparison of
  the HL-LHC and ILC prospects.   [DEL]

\item R-parity violating SUSY.  R-parity violation in the leptonic
  sector can lead to new resonant reactions such as $\ee\to \s \nu$
  and to new leptonic and hadronic decays of neutralinos.  Extensive
  searches were made for lepton R-parity violation at LEP (for
  example, \cite{Achard:2001ek,Abbiendi:2003rn,Abdallah:2003xc}), but
  there 
has been little work on the
  prospects for future $\ee$ colliders, even though existing LHC searches
  leave many possibilities open. A current study can be found in
  \cite{List:2013dga}.  A useful theory reference is
  \cite{AristizabalSierra:2004cog}.  [DEL]

\item New Higgs scalars -- pair production.  Models of new physics often
  contain new scalar bosons, perhaps from an extended Higgs sector.  
At an $\ee$ collider, the cross
  sections for pair-production of new scalar bosons are unambiguously
  predicted, depending only on  the masses and quantum numbers.   The
  situation is complementary to that of hadron colliders, where both
  production and decay rates depend on the detailed parameter
  choices.  To what extent can ILC fill in the gaps and exceptions in
  the search for new scalars left by the LHC?   Some studies of scalar
  pair production at $\ee$ colliders, for LEP, ILC, and higher-energy
  colliders,
 are given in
 \cite{Battaglia:2008nk,Battaglia:2012ia,Abbiendi:2013hk,Bahl:2020kwe}.
 [DEL]

\item New Higgs scalars -- $Z$ recoil.   Just as the Higgs boson appears as
  a resonance in the missing mass in $\ee\to Z^0 + X$, a new scalar can
  also appear as such a resonance, discoverable with couplings very small
  compared to the $hZZ$ coupling.  A current study, based on the use
  of $Z\to
  \mu^+\mu^-$ decays, is given in \cite{Wang:2020lkq}.  Can the use of hadronic
  decays of the $Z$ provide stronger constraints or sensitivity?
  [DEL]

\item New particles addressing the hierarchy problem.  The above
  questions relate to specific models that solve the gauge hierarchy
  problem.  But, in general, any solution to this problem requires
  some set of  new
  particles that appear in loop diagrams and cancel the ultraviolet
  divergences of the  SM.   These particles can be of many types,
  bosonic or fermionic (or both), colored or color-singlet. How
  general is 
  the ability of $\ee$ colliders to discover or exclude the various
  possibilities?  To what extent can $\ee$ measurements  test the ``naturalness sum
  rules''~\cite{Chen:2017dwb,Csaki:2018hyw} on masses and couplings 
that must be satisfied to cancel the ultraviolet divergences of the SM?  [TH]

\item   ILC response to LHC discovery.  For all of the models  discussed
  above,
  there is still ample phase space for the discovery of new particles
  at the HL-LHC.   Pick a particular example of a new particle that
  can be discovered, and discuss the additional information on the
  underlying model that can be obtained from $\ee$ experiments.  What 
$\ee$ center of mass energies are relevant?   (Remember that new
physics models typically have implications for modifications to Higgs boson couplings.)
A worked example can be found in~\cite{Fujii:2016raq}.  [TH]

 \item Relaxion.   The relaxion is a field postulated to solve the
   hierarchy problem by relaxing in the early universe to a ground
   state that picks out the 100~GeV mass scale~\cite{Graham:2015cka}.   It is
   possible for the relaxion to have a mass in the range of tens of
   GeV and to be observable through relaxion-Higgs mixing~\cite{Frugiuele:2018coc}.
   In this case, a measurement of the coupling of that particle to the
   Higgs boson can test the relaxion mechanism.  To what extent can an 
   $\ee$ collider confirm or refute this idea?  [TH]

\item Dark matter candidates -- effective Lagrangian approach.  
In the same way that dark matter
  production at the LHC can be parametrized by an effective Lagrangian
  with $q\bar q \chi \chi$ 4-fermion operators, dark matter production at the
  ILC can be parametrized by an effective Lagrangian with $e\bar e
  \chi \chi$ operators.   At the ILC, dark matter production is
  observed using initial-state photon emission.   For the various
  possible Lorentz structures, how close can we explore up  to  the limit
  $m_\chi = \half \sqrt{s}$?   Two useful current references are
  \cite{Habermehl:2020njb,Habermehl}.     [DEL]

\item Dark matter candidates -- photon-induced reactions.   A possible
  effective Lagrangian term that gives a portal to dark matter is the 
  coupling to photons: $F_{\mu\nu} F^{\mu\nu} \chi\chi $.  What limits
  can be obtained from photon-photon collisions at $\ee$ colliders?   [GEN]

\item  Dark matter candidates -- light mediators.  There are many
  scenarios in which the mediator that links the SM and dark matter
  sectors is  light -- 10s of GeV and below -- so that the effective
  Lagrangian description in which the mediator is integrated out is
  not valid.   It
  would be interesting to study the sensitivity in $\ee$ to a  particular model, perhaps one
  considered in the LHC studies~\cite{Boveia:2018yeb}, as the mediator mass
  varies from the 10 GeV to the 100 GeV region.   [DEL]

\item Dark matter candidates -- Higgs sector.   It is possible that
  extensions of the Higgs sector contain symmetries that would
  stabilize  a heavy neutral bosonic dark matter candidate.  Searches
  at $\ee$ colliders
  for new particles in a model of this type have been 
studied in \cite{Kalinowski:2018ylg,Zarnecki:2020swm}.  It is
interesting, then, to think about dark matter scenarios with
extended-Higgs-type signatures, and to compare the capabilities of ILC
and LHC  to discover these  models. [DEL]

\item  Dark matter candidates -- top quark sector.  It is possible
  that the most important process for dark matter production at high
  energies involves the radiation of the mediator  from a top quark, producing, for
  example, the final state $ t \bar t \ +$~(missing).  What constraints can
  one put on this dark matter production mechanism at $\ee$
  colliders?  [DEL]

\item Dark photons.  In the study of models with ``dark sectors'',
  there are benchmark ``visible'' and ``invisible'' $A'$ models,
  defined, for example, in \cite{Bjorken:2009mm,Berlin:2018bsc}.  The $A'$ is produced at an
  $\ee$ collider in $\ee\to A' + \gamma, Z$.   What is the sensitivity
  of the ILC 
  to these models  at the various ILC energy stages?  For the
  visible $A'$ models, compare the sensitivities from the various $A'$
  decay channels.  [DEL]

\item Axion-like particles (ALPs).  ALPs can be produced at $\ee$
  colliders in $\ee \to \mbox{ALP} + \gamma, Z, h$ and in $Z$ decays
  to $\mbox{ALP} + \gamma$, with ALP decay to $\gamma\gamma$.
   In some parameter regions, the ALP decay
  can be displaced.   A review of ALP searches at a variety of
  colliders can be found in \cite{Bauer:2018uxu}.    What is the sensitivity of ILC at its
  various energy stages?     [DEL]

\item  Long-lived particles.  New weakly coupled and long-lived
  particles could be pair-produced in $\ee$ collisions.   Scenarios
  with long-lived neutral particles are more difficult to constrain at
  the LHC  and provide an opportunity for discovery at the ILC.  It is
  interesting to analyze the separation of these signal events from
  physics and machine-induced backgrounds.   Some results for
  scenarios in which the new particles arise from Higgs decays can be
  found in \cite{Alipour-Fard:2018lsf}. Other sources, including
  photon-photon collisions and $WW$ fusion, may be more difficult to
  analyze. The ILC  machine and 
  detectors have a characteristic time structure with 5 or 10 bunch
  trains per second, each of total duration about 1 msec.  Between
  trains, the ILC detectors will be powered down to avoid the need for
  active cooling, an important element in the design of a precision detector
  with minimal material.  Does this affect the ability  to
  discover long-lived particles?  [FULL]

\end{enumerate}

\section{Questions about ILC fixed-target capabilities}

\begin{enumerate}

\item Beam-dump experiments.  ILC will produce intense, high-energy
  electron and positron beams that  will eventually pass through the
  ILC interaction region and be brought to beam dumps.  This would
  give a new opportunity to search for exotic dark-sector particles
  such as dark photons, millicharged particles and ALPs.   What is the
  reach of these experiments? A general orientation to fixed-target
  dark sector experiments can be found in \cite{Izaguirre:2013uxa}.
Experiments proposed for the near term are reviewed in the reports
\cite{Battaglieri:2017aum,Alemany:2019vsk}.     By how much does ILC extend their
  capabilities?  Some studies of this questions can be
  found in \cite{Kanemura:2015cxa,Sakaki:2020mqb}.    [GEN]

\item  Positron beams.  The availability of high-energy positron beams
  offers a novel type of fixed-target process, with positron
  annihilation on atomic electrons.  This process has been studied at
  low energy in \cite{Marsicano:2018glj}.  What capability is available with the ILC
  beams?  [GEN]
 
\end{enumerate}

\section{Questions about the theory of Higgs boson couplings}

\begin{enumerate}

\item  Higgs inverse problem.  Though predictions for deviations of
  Higgs couplings from the SM expectations have been computed in many
  models, there is still a poor understanding of the inverse problem:
  To what extent does the observation of a specific set of Higgs
  coupling deviations point to a specific class of BSM models.  How
  well can this relation be characterized?   Within the set of SUSY
  models, some of this relation is captured by general SUSY parameter
  fitting programs such as Fittino~\cite{Bechtle:2004pc},
  SFitter~\cite{Lafaye:2007vs}, and 
  MasterCode~\cite{Buchmueller:2010ai}.  [TH]

\item Higgs couplings from heavy SUSY.  There is a significant
  parameter space in 
 which SUSY models with very heavy superpartners (with squark masses
 at, say,
   5~TeV) can give rise to order 5\% deviations from the SM in the
  $hbb$ and $h\tau\tau$ couplings. Examples of such models are given
  in
\cite{Bahl:2020kwe,Cahill-Rowley:2014wba,Wells:2017vla}.   It would be interesting to
  explore more systematically the predictions for Higgs couplings
  in  SUSY models with masses too heavy to be
  discovered at the HL-LHC.  What distinct mechanisms can promote large
  deviations in the Higgs deviations, and what regions of SUSY
  parameter space are made accessible in this way?   [TH]

\item Higgs couplings to $WW$.   Although the SM Effective Field Theory allows couplings
  of the form $hW_{\mu\nu}W^{\mu\nu}$ in the leading order in BSM
  effects, such couplings are highly suppressed in BSM models with a
  weak-coupling description, including not only SUSY but also
  Randall-Sundrum and other extra-dimensional models.  Can these
  couplings arise in other types of composite Higgs models?  What
  would be the implications of the discovery of a coupling with this
  structure?   [TH]

\end{enumerate}

\section{Questions about SM Effective Field Theory interpretation of
  $\ee$ measurements}

\begin{enumerate}

\item Combination of ILC measurements.  The ILC physics program gains
  much of its power from the ability to combine measurements of Higgs
  processes with those of $\ee\to W^+W^-$ and precision electroweak
  observables in the context of
 SM Effective Field Theory (SMEFT)~\cite{Barklow:2017suo,deBlas:2019wgy}.
 From this point of view, what
  are the weak  links that require improved measurements or special
  attention?   [TH]

\item Combination of ILC and LHC.  Comparing the SMEFT analyses in 
\cite{Bambade:2019fyw} and \cite{deBlas:2019rxi}, the former group
includes only a few of the simplest LHC measurements 
 while the latter group proposes a more general fit
using the whole corpus of LHC Higgs data.   In some sense, this is a
trade-off between including more available information and reducing
the number of model assumptions.   Is there a more optimal way to combine ILC and
LHC information?     [TH]

\item Adequacy of SMEFT as a description of electroweak symmetry
  breaking.  It is possible that the 125 GeV Higgs boson is not the
  only source of $SU(2)\times U(1)$  symmetry breaking.  Additional
  sources of symmetry breaking may come from additional heavy Higgs
  multiplets or from the $SU(2)\times U(1)$-violating masses of heavy
  particles.  Integrating out these heavy particles yields an
  effective Lagrangian more general than SMEFT, called Higgs-Electroweak
  Effective Field Theory or
  HEFT~\cite{Buchalla:2013rka,Brivio:2013pma}.   This Lagrangian is
  non-analytic in the SM Higgs  doublet $\Phi$ or, equivalently, treats
  the Higgs scalar field and its associated Goldstone boson fields
  separately using an nonlinear realization of the 
symmetry~\cite{Chang:2019vez,Falkowski:2019tft}.    To what extent can
we test using
$\ee$ colliders 
whether  SMEFT is an adequate description or whether the additional
freedom in HEFT is required?  [TH]

\item Limitations of the SMEFT. In the SMEFT description of BSM
  models, the approximation  of considering only  dimension-6 operators 
  breaks down when the  BSM particles, assumed to be integrated out in
  SMEFT, have masses close to the Higgs boson and top quark masses or
  the process CM energy.
  Are there concrete models for which the use of the approximation of
  keeping dimension-6 terms only 
leads to an 
  incorrect or misleading interpretation of the data?    [TH]

\item  Higgs couplings in the presence of exotic decays.   Is there a
  formally correct way to include the possibility of both high-mass effects parametrized
  by SMEFT and of light particles giving new exotic Higgs decays in
  the same analysis of ILC data?  [TH]

\item Top quark in SMEFT.   The top quark appears in a large number of
  dimension-6 SMEFT operators.   There are 10 operators that
  modify top quark-vector boson form factors and another 10 operators
  that give 4-fermion contact interactions contributing to ILC
  observables.  At a fixed energy,  these two classes of
  operators affect observables in similar ways, so it is difficult to
  distinguish them.  The analysis
 \cite{Durieux:2019rbz} demonstrates
  that it is possible to determine these operator coefficients
  independently using ILC measurements at 500~GeV and 1~TeV.   This
  point merits further analysis.  What is the best way to obtain 
  independent determinations of these operator coefficients?      [TH]

\item Loop effects in SMEFT.  Almost all current $\ee$ SMEFT analyses
  treat SMEFT effects only at the leading order in perturbation
  theory.  Some exceptions are studies of loop effects of the top
  Yukawa
  coupling~\cite{Vryonidou:2018eyv,Durieux:2018ggn,Jung:2020uzh}  
 and the  Higgs
  self-coupling~\cite{DiVita:2017vrr}, in which only these specific 
 couplings are considered at 1-loop
  order.
Are there other important SMEFT loop effects that need to
  be taken into account in ILC analyses?    [TH]

\item Global SMEFT analysis.  Most SMEFT analyses in the literature
  discuss a specific subset of higher-dimension operators in isolation
  from the rest.   Separate analyses are done for precision
  electroweak interactions, $W$ trilinear couplings, Higgs couplings,
  top quark couplings, 4-fermion interactions, and the Higgs
  self-coupling. Studies that combine constraints from different
  sectors  (\eg,
\cite{Barklow:2017suo,DiVita:2017vrr,Falkowski:2016cxu,deBlas:2019wgy})
find interesting synergies and increased power.   At $\ee$ colliders,
it is surely easier than at hadron colliders to achieve a full global
analysis that constrains or, hopefully, overconstrains all possible
SMEFT contributions.  What is needed to achieve this goal?      [TH]

\end{enumerate}

\section{Contact information for the SiD and ILD detector groups}

For the most part, the questions listed above can be investigated
using the simplified simulation tools described in Sec.~3 above.
But some people would like to carry out deeper-level analyses,
investigating issues associated with detailed reconstruction
algorithms, detector optimization, and alternatives for detector
design.  For work at this level, there is a dedicated software framework
called ILCSoft~\cite{iLCSoft}.   Both SiD and ILD have constructed detailed detector
models based on this framework  and have used it to carry out
full-simulation analyses.  The analysis frameworks being used for CEPC
and CLIC analyses are also based on ILCSoft.   To learn this framework, we strongly
encourage you to join one of these existing collaborations.  Each has
a long list of low-level  reconstruction projects that  would be
suitable for contributions to Snowmass and would provide an accessible
starting point for developing your own ideas.

 Both SiD and ILD are offering  free guest memberships for participants in the
Snowmass study.  These will give access to the group resources and
technical support with the ILCSoft analysis packages.  To join ILD in
this way, please see~\cite{joinILD}.  To join SiD, please contact the spokespersons.

At this moment, participants in the four $\ee$ collider proposals are
collaborating in developing a common and  more modern software package
using the CERN ACTS framework~\cite{Ai:2019kze}.  But it is unlikely  that this package will 
be ready in time for Snowmass 2021 projects. 

To join the SiD group, please contact
\begin{itemize}
\item Spokespersons:   Andrew White (awhite@uta.edu), Marcel Stanitzki \\
  (marcel.stanitzki@desy.de)
\item Physics Coodinator:  Tim Barklow (timb@slac.stanford.edu)
\end{itemize}

To join the ILD group, please contact
\begin{itemize}
\item Spokesperson: Ties Behnke (ties.behnke@desy.de)
\item Physics Coodinators:  Keisuke Fujii (keisuke.fujii@kek.jp), \\
  Jenny List (jenny.list@desy.de)
\item Executive Team member from the US:  Graham Wilson (gwwilson@ku.edu)
\end{itemize}

\section{R\&D collaborations, and contact information for joining
  them}

Over the past few decades, a large number of groups have 
pursued extensive generic detector R\&D studies, applicable to any Linear
Collider (LC) detector concept.  This reseach has been carried
out both within the detector concept groups SiD, ILD, and CLICdp 
 and within collaborations such as  CALICE, LCTPC, and FCAL. The
 directions being pursued
 and an update of the most recent R\&D results are
 summarized in the ``Linear Collider Collaboration
Detector R\&D Report''~\cite{RDLreport}.   This document provides a
``snapshot'' 
entry point for new groups, to help them to learn
about the current landscape of the LC R\&D efforts and the areas where
they might be interested to contribute.  The latest update, dated
December 2018, was 
submitted as supplemental LCC input to the European Strategy Update.
 The next version will be released in July 2020.
Please contact the editors Jan Strube (jstrube@uoregon.edu) and Maxim
Titov (maxim.titov@cea.fr) for any questions about this document or for
further information about the LC R\&D program.

    There are a number of ``transversal''  R\&D collaborations dedicated to
    streamline effort and resources, handle new technologies, and
    match common components
 to on-going engineering developments or production.  
 Here is a list of the largest collaborations, with
 contact information and a summary of the questions that they address:
\begin{itemize}
\item CALICE — highly granular electromagnetic and hadron calorimetry 
\begin{itemize}
  \item \url{https://twiki.cern.ch/twiki/bin/view/CALICE/WebHome}
\item Spokesperson: Roman Poeschl (poeschl@lal.in2p3.fr)

\medskip

\item development of sensor technologies and electronics, mass
  production strategies and  system aspects
\item analysis of test beam data for detector performance,
  reconstruction algorithm;
\item    development and the study of hadronic shower physics
\end{itemize}

\item  FCAL — Highly compact and precise electromagnetic calorimeters
 for forward region of $\ee$ detectors
\begin{itemize}
\item \url{https://fcal.desy.de}
\item     Spokesperson: Wolfgang Lohmann (wolfgang.lohmann@desy.de)

 \medskip

\item development of ultrathin detector planes and dedicated electronics;
\item construction of prototypes;
\item performance studies using test-beam data;
\item study of the radiation tolerance of sensors
\end{itemize}

\item  LCTPC — Time Projection Chamber for a Linear Colider
\begin{itemize} 
\item \url{https://www.lctpc.org}        
\item    Spokesperson: Jochen Kaminski (kaminski@physik.uni-bonn.de)  
\item     Regional coordinator for the Americas: Alain Bellerive
  (alainb@physics.carleton.ca)

\medskip

\item construction/test-beams of TPC endplates with GEM, Micromegas, InGrid readout;
\item development/engineering challenges for ion blocking techniques (“gating”);
\item development of readout electronics, DAQ, cooling
\end{itemize}

\item SiDR\&D — silicon tracking and EM calorimetry for the SiD  detector concept
\begin{itemize}
\item          Spokesperson: Marty Breidenbach (mib@slac.stanford.edu)

 \medskip 

\item ultra-low-mass silicon sensors
\item high granularity electromagnetic calorimetry with small Moli\`ere radius
\end{itemize}
\end{itemize}

In addition, several groups world-wide are developing technologies for
precision vertex detectors, such as  CMOS MAPS, DEPFET, FPCCD, and
SOI.  Details on these studies can be found in the R\&D report.

\Acknowledgements

We are grateful to the members of the  LCC Physics and Detectors
Executive Board and the Americas Linear Collider Committee 
 and to Zhen Liu and Lian-Tao Wang  for contributions to this
document.   The work of the DESY group is supported by the Deutsche 
Forschungsgemeinschaft  under Germany’s Excellence Strategy, EXC 2121
``Quantum 
Universe'', grant 390833306. The work of the KEK group is supported in
part by the Japan Society for the Promotion of Science under the
Grants-in-Aid for Science Research 16H02173, and 16H02176. The work of
Graham Wilson is supported by the US National Science Foundation 
under award PHY-1913886.
 The work of James Wells  is supported 
in part by the  the US Department
of Energy,  contract  DE-SC0007859.  The work of Sunghoon Jung is
supported by the National Research Foundation of Korea under
 grant NRF-2017R1D1A1B03030820.  The work of the SLAC group is 
supported by the US Department
of Energy,  contract DE–AC02–76SF00515.  The work of Junping Tian
 is supported in
part by the Japan Society for the Promotion of Science under the
Grant-in-Aid for Science Research  15H02083.

\newpage

\appendix

\section{Event weight factors for various choices of electron and
  positron polarization}

\begin{table}[h!]
    \centering
	\begin{tabular}{|c|c|c|c|c|}
     \hline 
     & $e^-_L e^+_R$  & $e^-_R e^+_L$  & $e^-_L e^+_L$  &  $e^-_R e^+_R$ \\ 
     \hline 
                     & & & & \\
$(P_{e^-},P_{e^+})$  & $\frac{(1-P_{e^-})(1+P_{e^+})}{4}$ &
$\frac{(1+P_{e^-})(1-P_{e^+})}{4}$ 
&  $\frac{(1-P_{e^-})(1-P_{e^+})}{4}$ &  $\frac{(1+P_{e^-})(1+P_{e^+})}{4}$ \\    
                     & & & & \\
		\hline
$(-80\%,+30\%)$  & 0.585 & 0.035 & 0.315 & 0.065 \\               
$(+80\%,-30\%)$  & 0.035 & 0.585 & 0.065 & 0.315 \\               
$(-80\%,-30\%)$  & 0.315 & 0.065 & 0.585 & 0.035 \\               
$(+80\%,+30\%)$  & 0.065 & 0.315 & 0.035 & 0.585 \\               
		\hline
$(-80\%,+60\%)$  & 0.72  & 0.02  & 0.18  & 0.08  \\               
$(+80\%,-60\%)$  & 0.02  & 0.72  & 0.08  & 0.18  \\               
$(-80\%,-60\%)$  & 0.18  & 0.08  & 0.72  & 0.02  \\               
$(+80\%,+60\%)$  & 0.08  & 0.18  & 0.02  & 0.72  \\               
		\hline
	\end{tabular}
	\caption{Weight factors to reweight events with a given
          initial-state helicity for any general polarization, and
          specifically for the ILC baseline and possible upgrade values.}
	\label{tab:polweights}
\end{table}

\end{document}